\documentclass[a4paper]{article}
\usepackage{graphicx}

\begin{document}

\begin{center}
{\Large \bf The Possibility of New Physics in $pp$ Elastic Scattering at LHC}
\end{center}

\begin{center}
{
V. Uzhinsky\footnote{CERN, Geneva, Switzerland}$^,$
           \footnote{On leave of LIT, JINR, Dubna, Russia}}
\end{center}

\begin{center}
Abstract
\end{center}

\begin{center}
\begin{minipage}{12cm}
Modern models of high energy elastic hadron-hadron scattering predict an
oscillation character of differential cross sections at the LHC energy of 14 TeV
and at a sufficiently high momentum transfer. The Totem collaboration did not
see the oscillations at 7 TeV. According to some predictions, the oscillations
are weak at 7 TeV in the studied 4-momentum transfer range ($|t|<$ 2.5 GeV$^2$).
They may be beyond the range of the experiment. But a direct extension of the Totem collaboration
data on the pp-scattering at 7 TeV above $|t|\sim 2.5$ GeV$^2$ contradicts
previous measurements. Thus the collaboration can discover either
the oscillations at large $|t|$ or a change of the differential cross section
behavior in the high $|t|$ region ($|t|>$ 2.5 GeV$^2$).

\end{minipage}
\end{center}
For the first time, a collection of several theoretical model predictions of the
differential elastic $pp$-scattering cross sections at the LHC energy 14 TeV
\cite{Block14,Bourrely14,Desgrolard14,Petrov14,IslamMPLa24} was presented in
a paper by M.M. Islam et at \cite{IslamMPLa24}. It is shown in Fig. 1 as reproduced
from \cite{IslamMPLa24}. As seen, all of the models except one by M.M. Islam et al.
predict oscillations in the differential cross sections at $|t| > 2$ GeV$^2$.
As stated in that paper \cite{IslamMPLa24} -- "All these models predict visible
oscillations as well as much smaller cross sections than ours in the large $|t|$
region. Therefore, precise measurement of elastic $d\sigma/dt$ at large $|t|$ by
the TOTEM group will be able to distinguish between our model and the other models
and shed light on the dynamics of deep-elastic pp scattering".
\begin{figure}[cbth]
\includegraphics[width=160mm,height=100mm,clip]{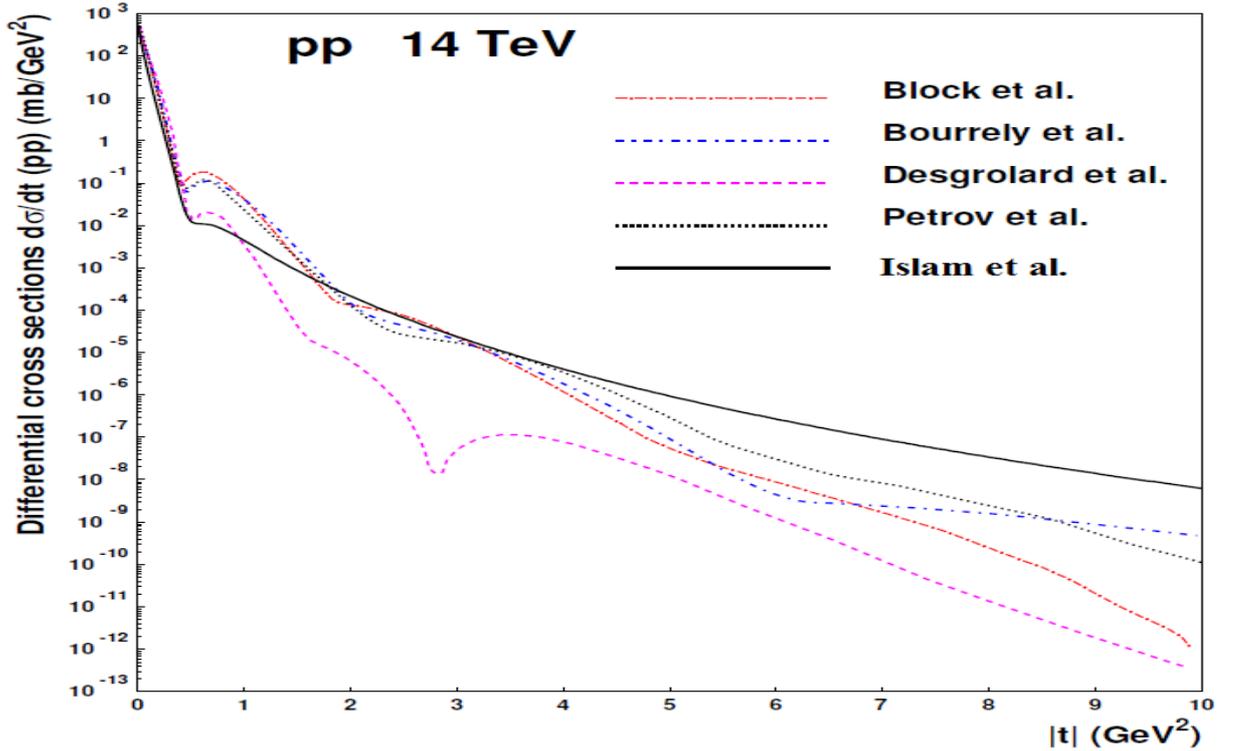}
\caption{Model predictions for $\sqrt{s}=14$ TeV. The figure is copied from
\protect{\cite{IslamMPLa24}}.}
\label{Fig1}
\end{figure}

A few more predictions are presented in \cite{ForwardPh}. They are not very different from
that given in Fig. 1.

The situation for predictions at 7 TeV center-of-mass energy is more complicated. Some
of these predictions \cite{Bourrely14,Block7,Jenkovzky7,Islam8} are shown in Fig. 2, in
which the oscillations do not appear clearly if it all. Perhaps with improved calculations,
the oscillations will appear.
\begin{figure}[cbth]
\includegraphics[width=160mm,height=80mm,clip]{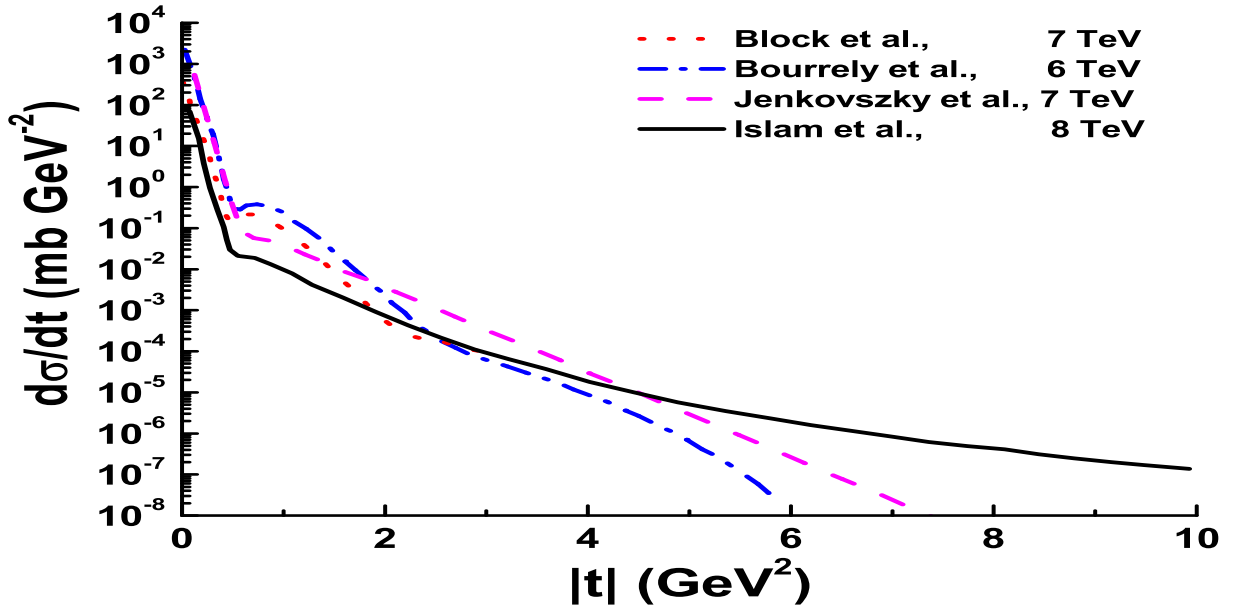}
\caption{Model predictions for $\sqrt{s}\sim 7$ TeV.}
\label{Fig2}
\end{figure}

In a paper \cite{OurPaper} a description of the Totem collaboration data \cite{Totem1,Totem2}
for elastic pp-scattering at $\sqrt{s}= 7$ TeV and 4-momentum transfer $|t|<2.5$ GeV$^2$ was
given. Extension of the calculations above $|t|=2.5$ GeV$^2$ shows the oscillations which are
due to the diffraction structure of the soft scattering. In that paper we proposed a unified,
systematic treatment of soft elastic scattering data starting from $P_{lab}\ >$ 10 GeV/c. In the description
of the high momentum transferred part we followed the approach of papers \cite{LandDonNP,Martynov}
which does not assume a complicated structure of the hard amplitude.

Recently, a new model of high energy elastic pp-scattering was published \cite{Selyugin}.
It describes the Totem data and predicts the oscillation also. So, the oscillations are expected
in most of the models. At the same time, in the last paper \cite{Dremin} damped oscillations are
predicted. Will the oscillations appear in differential elastic scattering?

The existing Totem data presented in Fig. 3 cannot give an answer. As seen, they are ended just in
the point where the oscillation can start. Maybe with extended measurements, the oscillations will
appear.
\begin{figure}[cbth]
\includegraphics[width=160mm,height=70mm,clip]{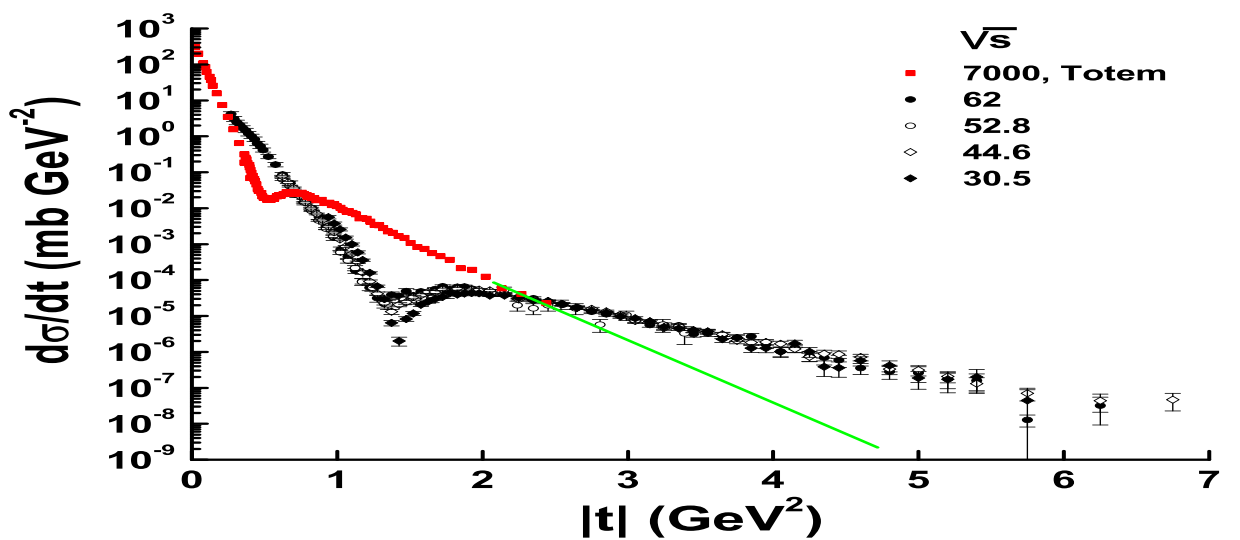}
\caption{$pp$ elastic scattering data at large momentum transfer. Points are experimental data
\protect{\cite{Totem1,Totem2},\cite{SS52_8},\cite{SS62},\cite{SS53}}.}
\label{Fig3}
\end{figure}

There is no doubt that a determination of the most reliable theoretical model is an important task
searching for the oscillations. More important, though, is a search for new phenomena, for which
a clear signature is predicted.

Previous low energy data on elastic $pp$ scattering are presented in Fig. 3 also. They
clearly contradict a direct extrapolation of the Totem data shown by the solid (green) line. We
look forward to a significant improvement in data at large $|t|$ in which the Totem collaboration
could discover the new phenomenon -- a change in the spectra at large $|t|$.

I believe that the Totem collaboration has all possibility to make a new discovery at the LHC --
to find the oscillation or the change.

The author is thankful to D.H. Wright, U. Wiedemann, V. Pozdniakov and A. Galoyan
for useful discussions. The author is also thankful to O.V. Selyugin for a sending of his calculations.


\begin{thebibliography}{1111}
\bibitem{Block14}M. M. Block, E. M. Gregores, F. Halzen and G. Pancheri, Phys. Rev. {\bf D60} (1999) 054024.

\bibitem{Bourrely14}C. Bourrely, J. Soffer and T.T. Wu, Eur. Phys. J. {\bf C28} (2003) 97.
\bibitem{Desgrolard14}P. Desgrolard, M. Giffon, E. Martynov and E. Predazzi, Eur. Phys. J. {\bf C16} (2000) 499.
\bibitem{Petrov14}V. Petrov, E. Predazzi and A. V. Prokudin, Eur. Phys. J. {\bf C28} (2003) 525.
\bibitem{IslamMPLa24}M.M. Islam, J. Kaspar and R.J. Luddy, Mod. Phys. Lett. {\bf A24} (2009) 485.

\bibitem{ForwardPh}R. Fiore et al., Int. J. Mod. Phys. {\bf A24} (2009) 2551.


\bibitem{Block7}M.M. Block and F. Halzen, Phys. Rev. {\bf D83} (2011) 077901.
\bibitem{Jenkovzky7}L.L. Jenkovzky A.I. Lengyel and D.I. Lontkovskyi, Int. J. Mod. Phys. {\bf A26} (2011) 4755.
\bibitem{Islam8}M.M. Islam, R.J. Luddy and A.V. Prokudin, Int. J. Mod. Phys. {\bf A21} (2006) 1.

\bibitem{OurPaper}V. Uzhinsky, A. Galoyan, 2011, arXiv:1111.4984 [hep-ph].
\bibitem{Totem1}TOTEM Collaboration, G. Antchev et al. Europhys. Lett. {\bf 95} (2011) 41001.
\bibitem{Totem2}TOTEM Collaboration, G. Antchev et al. Europhys. Lett. {\bf 96} (2011) 21002.

\bibitem{LandDonNP}A. Donnachie and P.V. Landshoff, Nucl. Phys. {B231} (1984) 189.
\bibitem{Martynov}E. Martynov, J.R. Cudell and A. Lengyel, 2005, arXiv:hep-ph/0509308.

\bibitem{Selyugin}O.V. Selyugin, 2012, arXiv:1201.4458 [hep-ph].
\bibitem{Dremin}I.M. Dremin and V.A. Nechitailo, 2012, arXiv:1202.2016 [hep-ph].

\bibitem{SS52_8}E. Nagy et al., Nucl. Phys. {\bf B150} (1979) 221.
\bibitem{SS62}U. Amaldi and K.R. Schubert, Nucl. Phys. {\bf B166} (1980) 301.
\bibitem{SS53}A. Breakstone et al., Phys. Rev. Lett. {\bf 54} (1985) 2180.


\end{thebibliography}
\end{document}